\documentclass[floatfix,twocolumn, longbibliography,pra,notitlepage, superscriptaddress,10pt, nofootinbib, groupedaddress]{revtex4-1}

\usepackage[utf8]{inputenc} 
\usepackage[T1]{fontenc}      

\usepackage[top=2cm,bottom=2cm,left=2cm,right=2cm]{geometry} 
\usepackage{verbatim}
\usepackage{amssymb,amsmath}
\usepackage{bbold}
\usepackage{graphicx}
\usepackage{float,bm}
\usepackage{hyperref}
\hypersetup{
  colorlinks   = true, 
  urlcolor     = blue, 
  linkcolor    = blue, 
  citecolor   = blue 
}
\usepackage{xfrac}
\usepackage{color}

\DeclareMathOperator{\Tr}{Tr}
\DeclareMathOperator{\Pf}{Pf}

\DeclareMathOperator{\re}{\mathrm{Re}}
\newcommand{ \drm }[0]{\mathrm{d}}

\newcommand{ \parO }[1]{\left(#1\right)}
\newcommand{ \matO }[1]{\left(\begin{matrix}#1 \end{matrix}\right)}
\newcommand{ \parS }[1]{\left[#1\right]}
\newcommand{ \parC }[1]{\left\{#1\right\}}

\newcommand{ \Om }[0]{\mathcal{O}}

\newcommand{ \ra }[0]{\rightarrow}
\newcommand{ \ket }[1]{\left| #1 \right \rangle}
\newcommand{ \bra }[1]{\left\langle #1 \right |}
\newcommand{ \braket }[2]{\left\langle #1 | #2 \right \rangle}

\newcommand{ \upa }[0]{\uparrow}
\newcommand{ \da }[0]{\downarrow}

\newcommand{ \id }[0]{\mathbb{1}}

\newcommand{ \lambb }[0]{\overline{\lambda}}
\newcommand{ \psib }[0]{\overline{\psi}}
\newcommand{ \etab }[0]{\overline{\eta}}

\allowdisplaybreaks
\newcommand{ \del }[0]{\partial}
\usepackage[most]{tcolorbox}
\usepackage{fancybox}
\usepackage[normalem]{ulem}

\begin{document}
\title{
	Quantum impurity models using superpositions of fermionic Gaussian states: \\Practical methods and applications
} 

\author{Samuel Boutin }
\author{Bela Bauer}
\affiliation{
    Station Q, Microsoft Corporation, Santa Barbara, California 93106 USA
    }
\date{\today}

\begin{abstract}
The coherent superposition of non-orthogonal fermionic Gaussian states has been shown to be an efficient approximation to the ground states of quantum impurity problems~[{Bravyi and Gosset, Comm. Math. Phys., \textbf{356} 451 (2017)}]. We present a practical approach for performing a variational calculation based on such states. Our method is based on approximate imaginary-time equations of motion that decouple the dynamics of each Gaussian state forming the ansatz. It is independent of the lattice connectivity of the model and the implementation is highly parallelizable. To benchmark our variational method, we calculate the spin-spin correlation function and R\'{e}nyi entanglement entropy of an Anderson impurity, allowing us to identify the screening cloud and compare to density matrix renormalization group calculations. Secondly, we study the screening cloud of the two-channel Kondo model, a problem difficult to tackle using existing numerical tools.
\end{abstract}

\maketitle

\section{Introduction}
Quantum impurity models -- systems of a few strongly interacting degrees of freedom coupled to a large bath of noninteracting fermions -- constitute an important class of problems in condensed matter physics. Despite the small number of interacting modes involved, this class of problems can exhibit rich many-body physics phenomena. The archetypical phenomenon is the Kondo effect, where even weak interactions can lead to strong non-perturbative corrections to the ground state~\cite{Wilson:1975ph}. Such models also appear as effective models in many embedding methods, such as dynamical mean-field theory~\cite{kotliar2006electronic}, that solve extended quantum many-body systems by approximately mapping them to quantum impurity problems.

Over the years, various numerical methods have been developed to tackle quantum impurity problems.
A particularly successful approach is Wilson's numerical renormalization group (NRG)~\cite{Wilson:1975ph} 
and its extensions~\cite{Bulla2008} which have allowed to study this class of problems in the thermodynamic limit. A related set of variational methods, based on the density matrix renormalization group (DMRG)~\cite{White:1992ao} has also been used extensively and compared to NRG~\cite{Weichselbaum2009,Saberi2008}. Finally, Quantum Monte Carlo methods have been successfully applied to systems where the sign-problem is mild, see e.g.~Refs.~\onlinecite{gull2011continuous,bertrand2019reconstructing}.
Despite these methods being very powerful, they each come with their limitations. For example, NRG is limited by an exponential scaling in the number of degrees of freedom of the impurity and the number of channels in the non-interacting bath; DMRG scales more favorably in the size of the impurity, but (since it does not exploit the non-interacting nature of the bath) retains an exponential scaling with the number of channels in the bath, rendering it very challenging to study, e.g., mesoscopic problems with several leads.
Quantum Monte Carlo methods, on the other hand, typically suffer from severe sign problems for multi-orbital systems.

A natural question is whether a well-chosen class of variational states could exploit the structure of quantum impurity models to circumvent the limitations of these established approaches. Recently, this question was affirmatively answered by proving that the ground states of quantum impurity problems can be approximated by a superposition of non-orthogonal fermionic Gaussian states~\cite{Bravyi2017}. This is obviously the case when the number of states in the superposition -- which we will refer to as the rank of the ansatz --  grows exponentially with the full system size, as the states then form a complete many-body basis.
More interestingly, the rigorous mathematical bounds of Ref.~\cite{Bravyi2017} demonstrate that the minimal rank to obtain a good approximation of the ground state scales only with the size of the impurity and the desired precision, while being independent of the size of the bath. This superposition of Gaussians (SGS) ansatz can be seen as a generalization of the generalized Hartree-Fock (GHF) method~\cite{bach1994generalized,Kraus_2010}, which aims to find the approximate ground state of a system using a variational minimization over the field of fermionic Gaussian states. While exact for a noninteracting system, GHF corresponds to a mean-field approximation for interacting systems. In the context of quantum chemistry, approaches related to our work are known as multi-component Hartree-Fock-Bogoliubov methods~\cite{Scuseria2011, Jimenez-Hoyos2012}.

Having chosen this ansatz, the challenge is to device practical algorithms to perform numerically efficient computations. Here, we focus on the problem of finding the lowest-energy state within the variational manifold. Multiple generic approaches exist for performing the energy minimization within a variational space, such as gradient descent and imaginary time evolution~\cite{Haegeman2011,Shi2017}. 
In particular, formal solutions for related ansatzes were originally developed under the name of resonating Hartree Fock~\cite{Fukutome1988,Fukutome1989,Tomita2004}. 
However, due to the large number of variational parameters in the SGS ansatz and the presence of several non-linear constraints, their numerical implementation can become prohibitively costly for large systems. 

In this work, we propose a simpler and numerically less costly path towards energy minimization within the variational manifold based on several key approximations to the imaginary time equations of motion. First, at each step we project the dynamics onto the subspace orthogonal to the one spanned by the current set of Gaussian states forming the SGS. Furthermore, we alternate the evolution of the coefficients of the coherent superposition of states and the (normalized) Gaussian states themselves. This allows us to decouple the equations of motion for each Gaussian state at each step in the evolution.
While the projection of these equation of motions onto the variational manifold does not exactly correspond to imaginary time evolution, we show that under this evolution, the energy is non-increasing. The variational state therefore converges to a local energy minimum within the manifold and can thus be used to study ground state properties of quantum models (we note that a guarantee on convergence to a global minimum typically cannot be given for variational algorithms).

To illustrate the power of the method, we apply it to two canonical impurity models: the single-impurity Anderson model~\cite{Affleck2008} and the two-channel Kondo effect~\cite{NozieresPh.1980}. The former has been studied using a variety of methods and is well-understood both analytically and numerically, thus allowing us to confirm the validity of our method. We find that using comparable computational resources, our method is able to achieve an error in the ground state energy that is about one order of magnitude better than DMRG. The two-channel Kondo model, on the other hand, is much more challenging to study numerically, and real-space correlation functions for fermionic leads had previously eluded numerical simulations. Instead, prior numerical studies have required either mapping to related problems in the same universality class~\cite{Alkurtass2016} or focusing on quantities that can be calculated using the local dynamics of the impurity~\cite{Mitchell2011}.

The remainder of this work is structured as follow. 
In Sec.~\ref{sec:probStatement},  we set the notation and describe the structure of the SGS ansatz.
In Sec.~\ref{sec:minProcedure}, we describe a generic minimization procedure for finding a variational approximation to the ground state and its numerical implementation.
In Sec.~\ref{sec:anderson1}, as a first demonstration of the method, we study the screening cloud of a single impurity Anderson model and benchmark our results using DMRG.
Finally, in Sec.~\ref{sec:MCK} we extend the calculations of the previous section by considering the two-channel Kondo model.

\section{Ansatz and problem structure}\label{sec:probStatement}
We start by describing the structure of fermionic quantum impurity models and of the SGS ansatz. We also introduce the covariance matrix formalism for Gaussian states which will be used throughout this work.

\subsection{Generic quantum impurity model}\label{sec:Model}
We consider a lattice model of $N$ fermionic degree of freedoms with Hamiltonian $H = H_2 + H_4$. We choose to work within a formalism of Majorana operators, noting that any fermion problem (both with and without particle-number conservation) can be rewritten in this form. The noninteracting part of the Hamiltonian is given by
\begin{equation}
    H_2 = i \sum_{k,l=1}^{2N} A_{k,l}c_k c_l,
\end{equation}
where $A$ is a real and skew-symmetric matrix, and $c_k=c_k^\dag$ is a set of $2N$ Majorana operators that obey standard anticommuation relations $\parC{c_j \, c_k} = 2 \delta_{j,k}$.
The interacting part of the Hamiltonian reads
\begin{equation}
    H_4 = \sum_{k,l,p,q}^{2N} U_{k,l,p,q} c_k c_l c_p c_q,
\end{equation}
where the rank-4 tensor $U$ is skew-symmetric with respect to the exchange of any neighboring indices. The interaction involves at most $M\ll N$ distinct Majorana operators making $U$ a sparse tensor. We remark that no assumption with regards to the lattice connectivity is made in our model definition. Although we focus here and below on quartic interaction terms, there is no fundamental limitation to including interaction terms involving a larger number of operators.

The ground state of the quadratic Hamiltonian $H_2$ will be a Slater determinant if $H_2$ conserves the number of particles or, more generally, a fermionic Gaussian state~\cite{Bravyi2004} in the case where the $U(1)$ symmetry is broken and only the parity of the number of particles is preserved. As Slater determinants constitute a subset of the fermionic Gaussian states, we will focus our discussion on Gaussian states. Although there might be a slight numerical overhead associated with working in this enlarged class of states, it has the advantage of naturally allowing the treatment of (mean-field) superconductivity.

\subsection{Covariance matrix formalism}\label{sec:CM-formalism}
Our description and usage of the covariance matrix formalism closely follows Ref.~\cite{Bravyi2017}. Any fermionic Gaussian state $\ket{\phi}$ obeys a Wick theorem and thus can be fully described by a covariance matrix (CM)
\begin{equation}
    \Gamma_{k,l} = \frac{-i}{2}\bra{\phi} [c_k \,,\, c_l] \ket{\phi},
    \label{eq:CMdef}
\end{equation}
where $k,l \in  1\dots 2N$. This matrix is real and skew-symmetric by construction. The expectation value of any product of Majorana operators can then be calculating as the Pfaffian of a submatrix of $\Gamma$.
For a normalized pure state the elements of the CM are subject to the constraint $\Gamma^2=-\id$.

Since the covariance matrices $\Gamma$ are matrices of expectation values, they are invariant under a gauge transformation $\ket{\phi} \ra e^{i \theta} \ket{\phi}$, with $\theta$ a real number. 
For calculations involving multiple Gaussian states $\ket{\phi_\mu}$ ($\mu=1,2, \dots)$, it is often necessary to fix this gauge freedom.
Following Ref.~\cite{Bravyi2017}, this can be achieved by choosing a reference state $\ket{\phi_0}$ and 
taking $\braket{\phi_0}{\phi_\mu}$ to be real and positive for all $\mu$.
Overlaps of Gaussian states and matrix elements can be obtained using the respective covariance matrices of the states~\cite{Bravyi2017}.

\subsection{Sum of Gaussian states ansatz}
Following Ref.~\cite{Bravyi2017}, the variational ansatz considered in this work is formulated as
\footnote{
        We follow the convention that Greek indices run over the labels of the states forming the ansatz  (i.e. $\alpha, \beta \dots \in 1\dots R$), while latin indices run over the Majorana operator labels (i.e. $k,l, \dots \in 1, \dots 2N$).
    }
\begin{equation}
    \ket{\psi} = \sum_{\mu=1}^{R} \lambda_\mu \ket{\phi_\mu},
    \label{eq:Ansatz}
\end{equation}
with $\parC{\ket{\phi_\mu}}$ a set of nonorthogonal Gaussian states, $\lambda_\mu$ complex scalar amplitude, and $R$ the rank of the ansatz.
This variational state is characterized by the  set of covariance matrices and amplitudes $\parC{\Gamma^\mu,\lambda_\mu | \mu =1 \dots R}$. 

This corresponds to $O(R N^2)$ variational parameters, subject to two normalization constraints. 
First,  the normalization of the variational state requires
\begin{equation}
    \sum_{\mu, \nu} \lambda_\mu \lambda_\nu^*  G_{\mu, \nu} =1,
\end{equation}
where we introduce $G_{\mu,\nu} = \braket{\phi_\mu}{\phi_\nu}$ the Gram matrix characterizing the overlap between states.
Second, we take each Gaussian state to be pure and normalized leading to the constraint on each covariance matrix $(\Gamma^\mu)^2 = -\id$. The sets of parameters obeying these constraints form the variational manifold.
Our aim is to find the state $\ket{\psi_0}$ that minimize the energy within this manifold.

Before turning to the full minimization problem, we first consider the following simpler problem.
Given a set of Gaussian states $\parC{\ket{\phi_\mu}}$ we wish to find the amplitudes $\parC{\lambda_\mu}$ which minimize the energy $E = \bra{\psi} H \ket{\psi}$.
These optimal amplitudes, leading to a normalized state with the lowest energy within the subspace, can be obtained by diagonalizing the Hamiltonian projected onto the subspace spanned by the set of Gaussian states. 
This leads to the generalized eigenvalue problem~\cite{Bravyi2017,Fukutome1988}
\begin{equation}
     h\lambda = E G\lambda
     \label{eq:eigLambda}
\end{equation}
where $h$ is the $R\times R$ matrix with elements $h_{\alpha,\beta} = \bra{\phi_\alpha}H \ket{\phi_\beta}$. In the case of orthogonal states $G=\id$ and this reduces to a regular eigenvalue problem.
This standard result will be at the core of the minimization approach introduced in Sec.~\ref{sec:minProcedure} as we will alternate between updating the amplitudes by solving Eq.~\eqref{eq:eigLambda} and updating the covariance matrices assuming fixed amplitudes.

\section{Projected equations of motion for energy minimization}\label{sec:minProcedure}
A generic approach for finding the ground state of a quantum system is imaginary time evolution. Starting from an initial state $\ket{\psi}$ and evolving according to the imaginary time Schrödinger equation 
\begin{equation}
    \del_\tau\ket{\psi} = -\parS{H- E_\psi}\ket{\psi},
    \label{eq:imtimeExact}
\end{equation}
where $E_\psi = \bra{\psi}H\ket{\psi}$,
allows to reach the ground state in the $\tau \ra \infty$ limit as long as the initial state has finite overlap with the ground state.
In the case where $\ket{\psi}$ is a variational state, the equations of motion must be projected back onto the part of the variational manifold orthogonal to $\ket{\psi}$ in order to best approximate the dynamics of the system and preserve the norm of the state.
Different methods were introduced to perform this projection~\cite{BROECKHOVE1988547}.
In the case of the time-dependent variational principle (TDVP), this projection requires the inversion of the Gram matrix of the tangent states (obtained by taking the derivative of $\ket{\psi}$ with respect to each variational parameter), which can be very large~\cite{Haegeman2011}. 
For the parametrization considered in this work, this does not appear to be a scalable approach as it would require repeated operations on matrices of dimension $RN^2\times RN^2$ leading to an $O(R^2N^6)$ numerical complexity.

In this section, we instead derive simplified projections of the imaginary time equation of motion for the SGS states.
Our approach can be understood as a parametrized energy descent.
We derive equations of motion for the covariance matrices $\Gamma^\mu$ as a function of an external parameter $s$ such that the energy $E(s) = \bra{\psi(s)} H \ket{\psi(s)}$
decreases monotically and converges to a local energy minima as $s\ra \infty$. 
As we do not pretend the followed approach to be sufficient to recover the system dynamics, we denote the evolution parameter as $s$ to distinguish the resulting equations from imaginary time (denoted $\tau$ above).

\subsection{Path of energy descent}
As eluded to at the end of Sec.~\ref{sec:probStatement}, at any instant $s$, one can separate the Hilbert space in two instantaneous subspaces, where one (referred to as "parallel subspace" below) is spanned by the set $\parC{\ket{\phi_\mu(s)}}$ of Gaussian states forming the ansatz, and the other is the orthogonal complement ("orthogonal subspace" below).
The energy minimization in the parallel subspace is easily performed by choosing amplitudes satisfying Eq.~\eqref{eq:eigLambda} and we will thus focus here on the orthogonal subspace assuming fixed amplitudes.

We consider as a starting point the imaginary time Schrödinger equation. However, instead of projecting on a subspace orthogonal to the instantaneous state $\ket{\psi (s)}$ leading to Eq.~\eqref{eq:imtimeExact}, we project onto a subspace orthogonal to the set $\parC{\ket{\phi_\mu(s)}}$ of Gaussian states
\begin{equation}
    \del_s \ket{\psi(s)} = - \Pi^\perp(s) H \ket{\psi(s)},
    \label{eq:projImEvol}
\end{equation}
where we have introduced the projector
\begin{equation}
    \Pi^\perp(s) = \id  - \sum_{\alpha, \beta} \ket{\phi_\alpha(s)} [G^{-1}(s)]_{\alpha, \beta} \bra{\phi_\beta(s)}.
\end{equation}
The inverse Gram matrix in the above equation ensures that $(\Pi^\perp)^2 = \Pi^\perp$ and is necessary since the Gaussian states considered are generally nonorthogonal (but linearly independent, insuring that $G$ is non-singular).
This equation of motion preserves the norm of the state since $\Pi^\perp(s) \ket{\psi(s)} =0 $ leading to $\del_s \braket{\psi(s)}{\psi(s)} =0$.

The projection in Eq.~\eqref{eq:projImEvol} constitutes a first approximation to the equations of motions and its justification is twofold.
First, assuming optimal state amplitudes, any dynamics lowering the energy should be orthogonal to the instantaneous parallel subspace. However, the resulting equation of motion is only approximate due to the additional implicit projection onto the Gaussian state manifold. This projection will become explicit through the use of the Wick theorem in Sec.~\ref{sec:CM-EOM}.
Second, the projection to the orthogonal subspace ensures that the different Gaussian states do not collapse into to a single state. Indeed, without it, all states would collapse to a mean-field approximation of the ground state independently.

Inserting the definition of the ansatz into the projected equation of motion Eq.~\eqref{eq:projImEvol}, we obtain
\begin{equation}
    \sum_\mu \lambda_\mu  \del_s \ket{\phi_\mu} = 
    -\sum_\mu \lambda_\mu \Pi^\perp(s) H \ket{\phi_\mu(s)}.
    \label{eq:projImEvol2}
\end{equation}
In order to move forward, we decouple the equation of motions of the different Gaussian states by postulating that there exists an effective Hamiltonian $B_\mu$ such that 
\begin{equation}
    \del_s \ket{\phi_\mu} = - B_\mu \ket{\phi_\mu},
    \label{eq:PhiMu}
\end{equation}
and which satisfies Eq.~\eqref{eq:projImEvol2}.
Upon inspection, one can find that
\begin{equation}
    B_\mu = c_\mu
    \Pi^\perp H \ket{\psi} \bra{\psi}  + h.c.
    \label{eq:Bmu}
\end{equation}
decouples the equations, with $c_\mu$ a complex scalar. Taking $c_\mu  = [\lambda_\mu \tilde{\lambda}_\mu^*]^{-1}$, with the renormalized amplitude $\tilde{\lambda}_\mu = \braket{\phi_\mu }{\psi}$,
satisfies Eq.~\eqref{eq:projImEvol2} confirming the validity of the decoupling scheme. 
Although this is not a unique choice, this approach ensures a monotonic decrease of energy, i.e. $\del_s \bra{\psi} H \ket{\psi} \leq 0$, upon simultaneous integration of Eq.\eqref{eq:PhiMu} for all states.

More generally, ensuring a path of energy descent upon the evolution of a given state $\ket{\phi_\mu}$
leads to the constraint $\re \parS{ c_\mu \tilde{\lambda}_\mu^* \lambda_\mu} >0$.
Hence, taking $c_\mu = \tilde{\lambda}_\mu \lambda_\mu^* / |\tilde{\lambda}_\mu \lambda_\mu|$, possibly up to a real and positive multiplicative factor for each state, 
is sufficient to decrease the energy of the variational state.

This decoupling scheme is the second major approximation to the equations of motion used in this work. While exact for a generic many-body basis, the decoupling is approximate in the case where the evolution is projected on a constrained variational manifold.
As is shown numerically in Secs.~\ref{sec:anderson1} and \ref{sec:MCK}, the optimization nevertheless converges towards the ground state.

\subsection{Equation of motion for the covariance matrices}\label{sec:CM-EOM}
In order for the above equations of motion to be useful, there must exist an efficient numerical implementation of them. We now derive the counterpart of Eq.~\eqref{eq:PhiMu} in the covariance matrix (CM) formalism.

Taking the derivative of the CM, as defined in Eq.~\eqref{eq:CMdef}, with respect to the evolution parameter $s$ we obtain (taking $k\neq l$)
\begin{align}
    \del_s \Gamma^\mu_{k,l} &= 
    -\bra{\phi_\mu} i c_k c_l \ket{\del_s  \phi_\mu}
    +c.c. , 
\end{align}
where $c.c.$ denotes the complex conjugate. Inserting Eq.~\eqref{eq:PhiMu} and the definition of $\Pi^\perp$ one obtains after some algebra
\begin{align}
    \begin{split}
    \del_s \Gamma^\mu_{k,l} &=
        i\sum_{\beta} \lambda_\beta \tilde{\lambda}_\mu^* c_\mu
     \parS{ 
         \bra{\phi_\mu}  c_k c_l  H \ket{\phi_\beta}  
         \vphantom{\sum_a}\right. \\  & \left. \qquad
         -\sum_{\gamma}
         \bra{\phi_\mu} c_k c_l \ket{\phi_\gamma}
         [G^{-1} h]_{\gamma, \beta}  
     }
     +c.c.
    \end{split}
    \label{eq:EOM-Gamma2}
\end{align}
In the special case of a rank 1 ansatz, Eq.~\eqref{eq:EOM-Gamma2} falls back onto the equation of motion for imaginary time evolution in the GHF approximation as derived for example in Ref.~\cite{Kraus_2010}.

We now specialize to the quantum impurity model with quartic interacting defined in Sec.~\ref{sec:Model}.
In order to rewrite the differential equation purely as a matrix equation,
we introduce the complex skew-symmetric matrices~\cite{Bravyi2017}
\begin{equation}
    \Delta^{\alpha, \beta} = \parS{i(\Gamma^\alpha - \Gamma^\beta)-2\id}(\Gamma^\alpha + \Gamma^\beta)^{-1},
    \label{eq:def-delta-matrix}
\end{equation}
which allows to compute easily matrix elements between different Gaussian states
$\bra{\phi_\beta}c_k c_l\ket{\phi_\alpha} = iG_{\beta,\alpha} \Delta_{k,l}^{\alpha,\beta}$ ($k\neq l)$. More generally matrix elements involving $n$ distinct Majorana operators are proportional to the Pfaffian of an $n \times n$ submatrix of $\Delta^{\alpha , \beta}$.
Similarly to GHF, we introduce generalized Fock matrices
\begin{equation}
    F^{\alpha, \beta}_{k,l} = A_{k,l} +6 \sum_{m,n} U_{k,l,m,n} \Delta_{m,n}^{\alpha, \beta},
\end{equation}
which for $\alpha=\beta$ ($\Delta^{\alpha, \alpha} = \Gamma^\alpha$) falls back on the standard Fock matrix for GHF~\cite{Kraus_2010}.

With these definition, the matrix differential equation for each CM takes the form
\begin{align}
        \del_s \Gamma^\mu &=
            \sum_{\beta} \lambda_\beta \tilde{\lambda}_\mu c_\mu
         \parS{ 
            i2G_{\mu, \beta}  [\Delta^{\beta, \mu}, F^{\beta, \mu}]
                \vphantom{\sum_a}\right. \nonumber
                \\  & \left. \qquad
                +
                2G_{\mu, \beta} \parO{
                \Delta^{\beta, \mu} F^{\beta, \mu} \Delta^{\beta, \mu} + F^{\beta, \mu}}
                \label{eq:EOM-Gamma3}
                \right. \\  & \left.\qquad
                -h_{\mu,\beta} \Delta^{\beta,\mu}
             +\sum_{\gamma}
             G_{\mu, \gamma}[G^{-1} h]_{\gamma, \beta} \Delta^{\gamma,\mu}
         }
         +c.c., \nonumber
\end{align}
where the first and second line are respectively reminiscent of the equation of motion for real and imaginary time evolution of a Gaussian state~\cite{Kraus_2010}. The third line ensures normalization and cancels in the case $R=1$. We have verified numerically that Eq.~\eqref{eq:EOM-Gamma3} preserves the norm and the purity of the state which requires
\begin{equation}
    (\del_s \Gamma^\mu)\Gamma^\mu + \Gamma^\mu (\del_s \Gamma^\mu) = 0,  
\end{equation}
due to the constraint $\Gamma^2=-\id$ for a pure state.
Equation~\eqref{eq:EOM-Gamma3} constitutes one of the main results of this work. 

\subsection{Numerical implementation}\label{sec:num}
In order to find a good approximation of the ground state, we alternately evolve the covariance matrices for a small step $s \ra s+\delta s$ and update the amplitudes by solving Eq.~\eqref{eq:eigLambda}. 
The computational complexity of each of these steps is $O(R^2 N^3)$ (assuming $R\ll N$). Since in each step, each Gaussian state can be evolved separately, one can easily parallelize over them  and achieve a significant speedup. 
Appendix~\ref{app:numerical_details} presents additional details on the numerical implementation. 

As an aside, we note that considering $P$ interaction terms with $m> 4$ Majorana operators could, in the worst case, lead to additional operations of computational complexity  $O(P R^2 N^2 m^3)$. However, in the most relevant case where $m,P\ll N$,
this shouldn't affect the scaling of the overall computational complexity of the method.
We note that the above estimate is an upper bound on the complexity of the right-and-side of Eq.~\eqref{eq:EOM-Gamma2} separately for each term of weight $m$ and each matrix element. Taking advantage of the skew-symmetric matrix structure of the CM might allow further algorithmic improvement. For example, in the case of $P=N^2$ quadratic operators ($m=2$) the naive estimate lead to complexity $O(R^2N^4)$ while the matrix formulation of  Eq.~\eqref{eq:EOM-Gamma3} is reduced to $O(R^2N^3)$.

In order to reduce the risk of converging to a local energy minima, we use in the remainder of this work the following strategy. Starting from the mean field (GHF) solution we gradually increase the rank of the ansatz by 1 after every $n$ iterations. The additional Gaussian state is chosen by applying a random special orthogonal transformation $Q$ to one of the current states of the ansatz with a high amplitude. 
For $n$ sufficiently large and $||Q-\id||_F$ sufficiently small we have found numerically this approach to converge towards a good approximation of the ground state. 
This approach is needed due to our choice of optimal amplitudes at each step of the algorithm. If all states are added at once, there is a strong risk to converge to a solution where some of the amplitudes vanish, effectively leading to an approximate ground state of a lower rank.

\section{Anderson impurity and the screening cloud}\label{sec:anderson1}
As a benchmark of the method, we study real-space properties of the single impurity Anderson model (SIAM)~\cite{Affleck2008}. This well-known model was previously studied numerically using a multitude of methods including NRG~\cite{PhysRevB.75.041307} and DMRG~\cite{Holzner2009,Nuss2015}. We will use the latter method to benchmark our results and confirm their validity.

We consider a 1D lattice where the SIAM Hamiltonian takes the form $H = H_0 + H_I$, with
\begin{align}
    H_0 &= -t \sum_{r=2}^{L-1}  \parO{d^\dag_{r-1}d_r + h.c.} 
     - \mu \sum_{r=1}^{L-1} n_r,
\end{align}
describing a bath of $L-1$ free fermions with hopping parameter $t$ and chemical potential $\mu$, and the impurity Hamiltonian
\begin{align}
    H_I = - t' \parO{d^\dag_0 d_1 + h.c.} 
    + \epsilon_d n_0 + Un_{0,\upa} n_{0,\da}.
\end{align}
Here, $d^\dagger$ are fermionic creation operators, $n_r = d^\dag_r d_r =\sum_{\sigma} d^\dag_{r,\sigma} d_{r,\sigma}$ is the number operator, and when omitted, spin indices are summed.
The hopping between the impurity (site 0) and the first site of the lead is  $t'$ and  $U>0$ is a repulsive interaction. We focus on the particle-hole symmetric point at half-filling where $\epsilon_d=-U/2$ and $\mu=0$. 

In the weak-coupling limit $t'<U$, charge fluctuations are suppressed on the impurity site creating an effective spin $1/2$ impurity.
The Hamiltonian then maps to the Kondo model with effective coupling strength $J = 8\Gamma/U$, where $\Gamma = (t'^2)/t$ is the broadening of the impurity energy level by the leads~\cite{Schrieffer1966}. The ground state is then a singlet state formed by the impurity spin and the collective spin of a delocalized so-called \emph{Kondo cloud} of electrons from the leads. 
One can associate to the cloud a length scale 
$\xi_K  \sim v_F / T_K$ where $v_F$ is the Fermi velocity and $T_K$ is the Kondo temperature~\cite{affleck2010kondo}.

\begin{figure}[tb]
    \centering
    \includegraphics[width=0.49\textwidth]{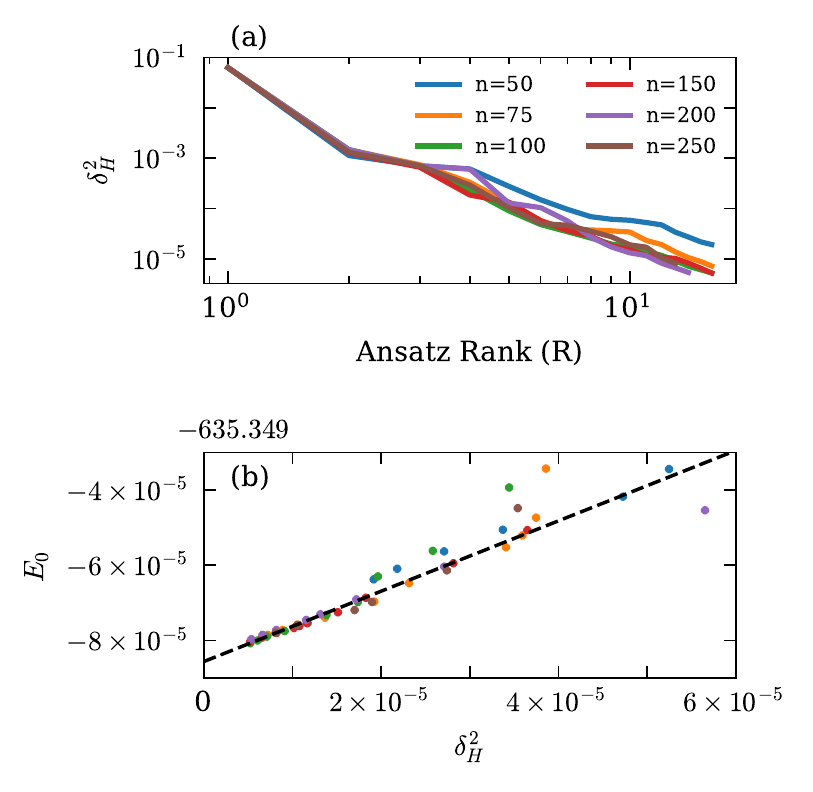}
    \caption{(a) 
        Convergence of the variance of the Hamiltonian $\delta_H^2$ (Eq.~\ref{eq:residual}) as a function 
        of the number of iterations $n$ per Gaussian state (see main text).
        (b) 
        Extrapolation using a linear fit (dashed black curve) of points with $\delta_H^2<2\times 10^{-5}$. 
        The obtained intercept, 
        $E^* = -635.3490861$,
        estimates the energy of the ground state up to a precision given by the standard error of the fit $\sigma_{E^*} = 4 \times 10^{-7}$.
        We consider a system of length $L=500$ at the particle-hole symmetric point ($\mu=0$, $\epsilon_d=-U/2$) with interaction strength $U=t=1$ and coupling $U/\Gamma=5$.
    }
    \label{fig:convergence}
\end{figure}
\begin{figure}[t]
    \centering
    \includegraphics[width=0.49\textwidth]{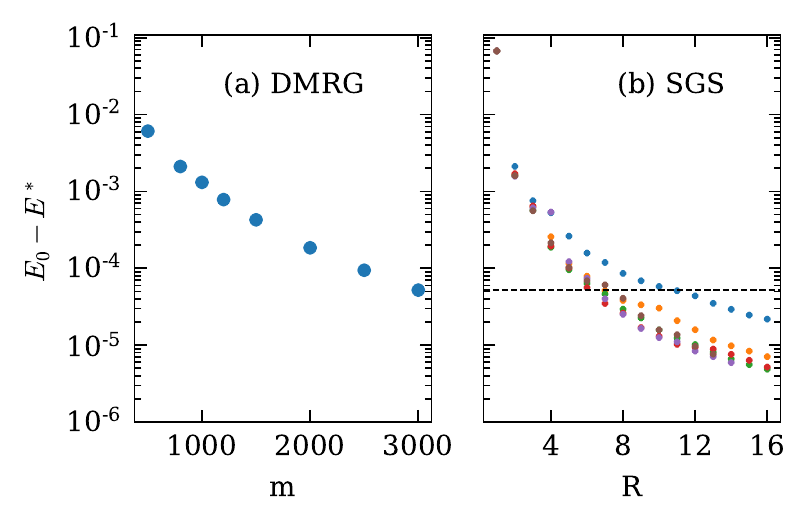}
    \caption{
        Convergence of the ground state energy obtained from (a) DMRG  and (b) the SGS ansatz respectively as a function of the bond dimension $m$ and the rank of the SGS ansatz. Dashed black horizontal line is the lowest energy obtained using DMRG.
        The ground state energy estimate $E^*$ is obtained from the extrapolation of Fig.~\ref{fig:convergence}(b). Parameters and legend correspond to those of Fig.~\ref{fig:convergence}.
    }
    \label{fig:EnergyDMRG}
\end{figure}

\subsection{Convergence and comparison to DMRG}
As a first test of the convergence of the method, we compute the variance of the Hamiltonian in the variational ground state,
\begin{equation}
    \delta_H^2 =  \bra{\psi} \parO{H -\langle H\rangle}^2 \ket{\psi},
    \label{eq:residual}
\end{equation}
where $\langle H\rangle = \bra{\psi}H \ket{\psi}$. 
Figure~\ref{fig:convergence}(a) presents $\delta_H^2$  as a function of the ansatz rank and for different number of iterations of the variational minimization (see Sec.~\ref{sec:num}) for a system size $L=500$.
The variance $\delta_H^2$ should converge to zero as the variational state converges to an eigenstate of the system ($\delta_H^2=0$ for any eigenstate of $H$). 
We estimate in Fig.~\ref{fig:convergence}(b) the converged ground state energy $E^*$ using a linear extrapolation of the variational energy as $\delta_H^2 \rightarrow 0$ for $\delta_H^2 \lesssim 2\times 10^{-5}$.

As a second test of the validity of our variational SGS state results,  
Fig.~\ref{fig:EnergyDMRG} compares the energies obtained using the SGS ansatz to energies obtained using state-of-the-art DMRG simulations~\cite{ITensor}. In both cases, we plot the results as a function of the refinement parameter of the ansatz, which is $R$ for the SGS ansatz and the bond dimension $m$ for DMRG. For the parameters considered, which were chosen to span similar computation times, the SGS-based method reaches lower energies than DMRG.

\begin{figure}[t]
    \centering
    \includegraphics[width=0.49\textwidth]{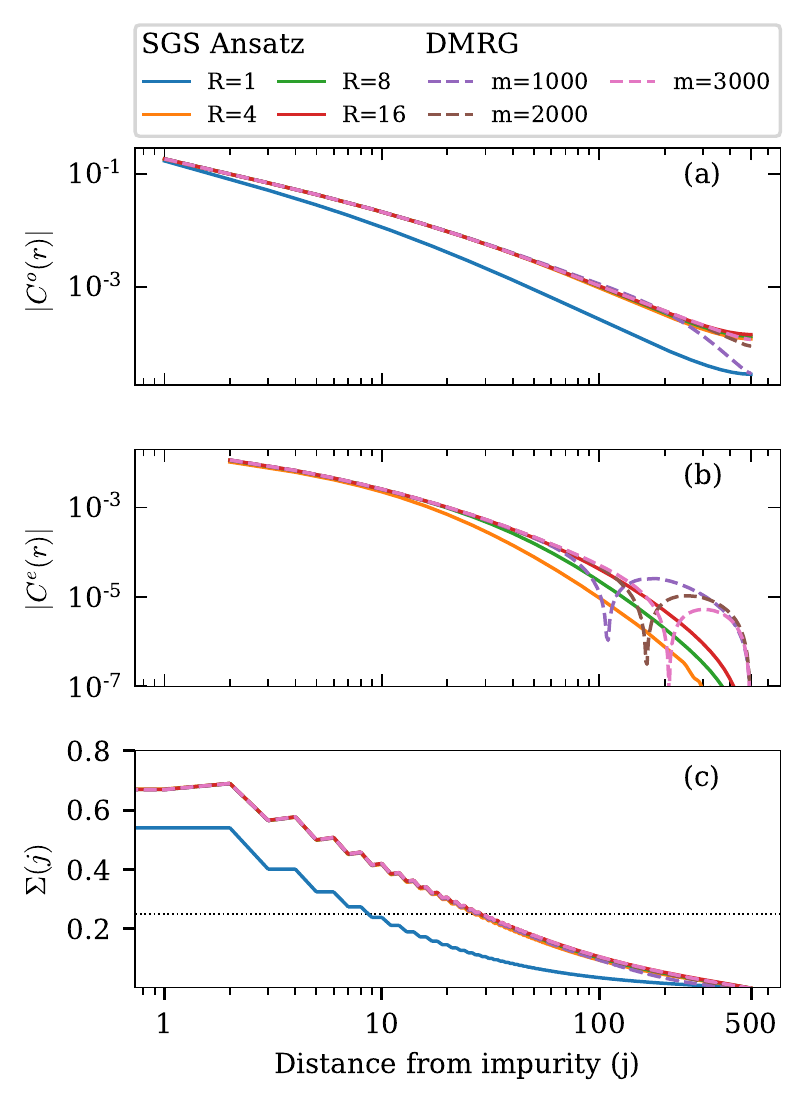}
    \caption{Comparison of the spin-spin correlation function $C(r)$ calculated using DMRG (dashed curves) and the SGS ansatz (solid curves).
    (a) Odd sites and (b) even sites. The mean-field solution ($R=1$, blue solid curve) is absent from panel (b) as $C^e(r)$ is identically zero up to numerical precision.
    (c) Integrated correlation function as defined in Eq.~\eqref{eq:integrated_correlation}.
    See Fig.~\ref{fig:convergence} for parameters.
    }
    \label{fig:SpinSpinDMRG_SGS}
\end{figure}
\subsection{Screening cloud - Spin-spin correlation}\label{sec:spinspin}
To better check the validity of the variational solution, we calculate the ground-state equal-time spin-spin correlation function between the impurity and the sites along the lead
\begin{equation}
    C(r) = \langle \mathbf{S}(0)\cdot \mathbf{S}(r) \rangle
    \label{eq:Cr}
\end{equation}
where the spin operator is $ \mathbf{S}(r) = \frac{1}{2}\sum_{s,s'} d^\dag_{r,s} \boldsymbol{\sigma}_{s,s'} d_{r,s'}$.

From a low-energy expansion of the bath operators~\cite{Barzykin1998} away from the impurity ($k_F r\gg 1$, with $k_F$ the Fermi wavevector), the correlation function is expected to be the sum of uniformly decaying and oscillating functions 
\begin{equation}
    C(r) = C_{U} (r) +C_{2k_F}(r)\cos (2k_F r)
\end{equation}
where, at zero temperature, $C_{U} (r)$ and $C_{2k_F}(r)$ are smoothly decaying functions.
As we focus on a half-filled lattice where $k_F = \pi/2a$ (with $a$ the lattice constant), the correlation function is the sum of  uniform and staggered 
($\cos (2k_Fr) \rightarrow (-1)^r$ ) contributions.

Figure~\ref{fig:SpinSpinDMRG_SGS} compares the correlation functions computed using the approximate ground states obtained with SGS ansatz and DMRG.
To plot more easily the highly oscillating function $C(r)$, we introduce the correlation function on the even (e) and odd (o) sites, denoted as $C^e(r)$ and $C^o(r)$, respectively.
For the odd sites (Fig.~\ref{fig:SpinSpinDMRG_SGS}(a)), the uniform and staggered parts of the correlation function are both negative, leading to a larger amplitude than the even sites (Fig.~\ref{fig:SpinSpinDMRG_SGS}(b)) where the contributions have opposite signs.
Focusing first on the odd sites, the mean-field solution ($R=1$, solid blue curve) differ qualitatively from the higher-precision DMRG results as expected. Modestly increasing the rank of the SGS ansatz, we recover the same behavior as high-precision DMRG. Small discrepancies between the two methods are observed far from the impurity where the amplitude of the correlation function is small. 
These differences are consistent with the expected precision of both methods and
we expect that further increasing the values of $m$ and $R$ would reduce these differences.
Similarly, for the even sites, both methods agree close to the impurity. Notably, the DMRG results shows a change of sign of $C^e(r)$ far from the impurity. This effect has been previously observed in other DMRG studies of the SIAM~\cite{Holzner2009,Nuss2015} and appears to disappear as the bond dimension of the MPS is increased. No such effect is observed for the SGS state, thus suggesting that this feature in the DMRG results is an artifact of the truncation of the MPS bond dimension.
 
As a final comparison between methods, we consider in Fig.~\ref{fig:SpinSpinDMRG_SGS}(c) the integrated correlation function
\begin{equation}
    \Sigma(x) = 1+ \sum_{r=1}^x \frac{C(r)}{C(0)},
    \label{eq:integrated_correlation}
\end{equation}
which allows characterization of the screening cloud of the impurity~\cite{Holzner2009}.
In the ground state, the total spin $\mathbf{S}_{\rm tot}^2$ is expected to be zero for even $L$, leading to the sum rule $\Sigma(L-1) = 0$.
While this sum rule is not explicitly enforced, it is approximately respected and the violation converges towards zero as the rank of the ansatz is increased ($\Sigma(L-1) \approx 2 \times 10^{-6}$ for $R=16$).

\begin{figure}[t]
    \centering
    \includegraphics[width=0.49\textwidth]{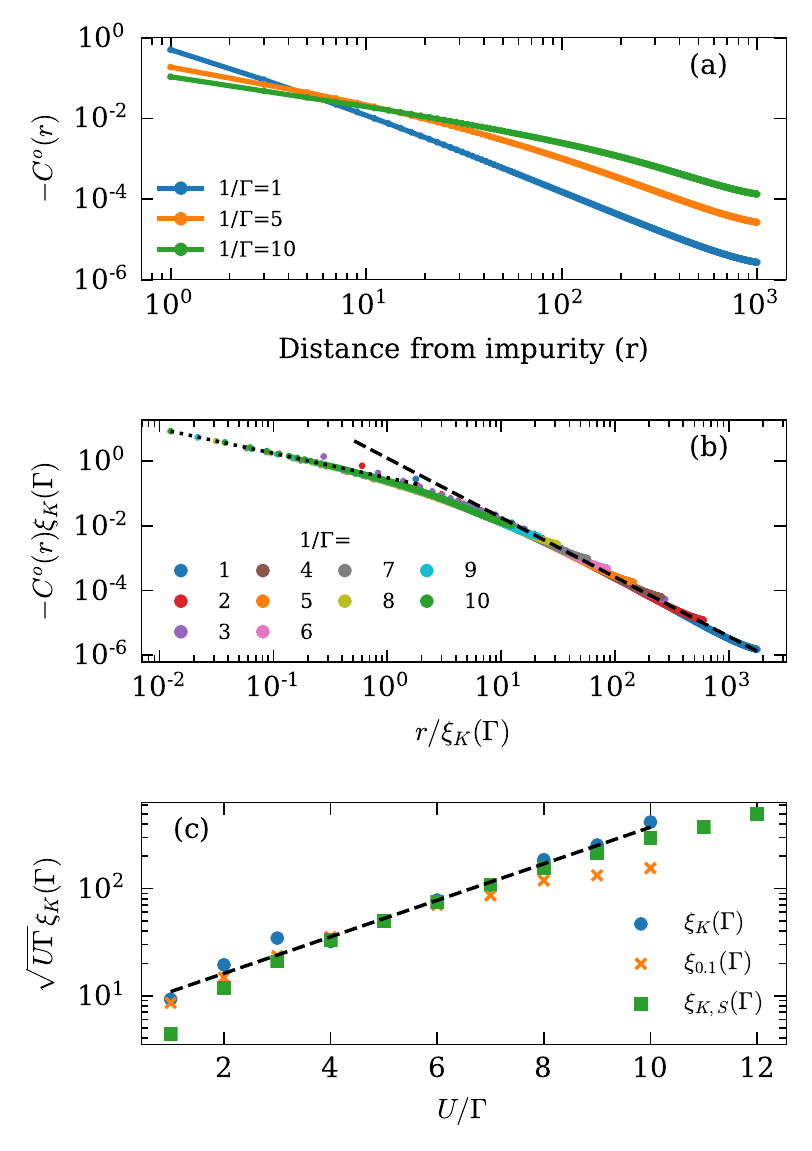}
    \caption{
    (a) 
    Spin-spin correlation function of a SIAM with $L=1000$ (ansatz Rank $R=14$) 
    on odd sites
    for coupling strengths $\Gamma=1.0$ (dark blue curves) and $\Gamma=0.1$ (light cyan curves).
    (b) Scaling collapse of $C^o(r)$. 
    Dashed (dotted) black curve is a fit of the form $A(r/\xi_K)^{-\nu}$ to the data with $r>10\xi_K$ ($r<\xi_K/2$), with exponent $\nu \approx 1.84$ ($\nu \approx 0.76$).
    (c) Length scales $\xi_K$ extracted from the scaling collapse of panel (b) (blue disks), $\xi_{0.1}$ extracted from the integrated correlation function (orange crosses, cf Eq.~\ref{eq:xic}) and $\xi_{K,S}$ extracted from the scaling collapse of the impurity entropy (green squares).
    See Sec.~\ref{subsec:S2SIAM}  for a discussion of the impurity entropy.
    Black dashed line is a fit of the expected functional form $A\exp\parS{\pi U/8\Gamma}$ to $\sqrt{U\Gamma}\xi_K(\Gamma)$. 
    As $\xi_K(\Gamma)$, $\xi_{0.1}(\Gamma)$  and $\xi_{K,S}$ 
    are related up to a scaling parameters,
    we set $\xi_K(1/5) =\xi_{0.1}(1/5)=\xi_{S,K}(1/5)$. 
    To reduce finite-size effect we append to the end of the bath an additional $L_W=20$ sites with an exponentially decreasing hopping parameter $t_n = \Lambda^{-n}t$ with $\Lambda=1.2$ and $t=1$.
    }
    \label{fig:K1}
\end{figure}
To conclude this section, we use the SGS ansatz to study the screening cloud of the Anderson impurity for a large system of $L=1000$ sites and different coupling strengths $\Gamma = (t')^2/t$. Figure~\ref{fig:K1}(a) shows $-C^{o}(r)$ for three different couplings between the impurity and the bath. In the Kondo regime, the correlation function $C^o(r,\Gamma)$ is expected to collapse to a universal function $\tilde{C}(r/\xi_K)$ through the relation $\tilde{C}(r/\xi_K) = \xi_K(\Gamma)C^o(r,\Gamma)$ with $\xi_K(\Gamma)$ a coupling-dependent length scale. 
Figure~\ref{fig:K1}(b) present the scaling collapse of ten different curves for the odd distances from the impurity $C^o(r)$.\footnote{The scaling collapse is obtained by minimizing the square of the distance on a log-log scale  between the numerical data points and a polynomial fit to $\xi_K(\Gamma) C^o(r)$ as a function of $r/\xi_K(\Gamma)$ for all $\Gamma$ values considered.} 
As expected from previous numerical studies~\cite{PhysRevB.75.041307} and analytical calculations~\cite{Barzykin1998}
the scaled correlation function decays following a power law $ \tilde{C}(r/\xi_K) \propto (r/\xi_K)^{-\nu}$ with a crossover in the exponent $\nu$ near $r/\xi_K=1$.
From power-law fits, we extract the exponents $\nu\sim0.76$ for $r\ll \xi_K$ and $\nu\sim1.84$ for $r\gg \xi_K$.
Given the sensitivity of the results of the fit on the range of parameters considered, this results are consistent with the previously established scenarios of a crossover from $\nu=1$ to $\nu=2$ near $r\sim \xi_K$.

Finally, Fig.~\ref{fig:K1}(c) presents the impurity screening length scale as extracted using three different methods. 
First, the parameters $\xi_K(\Gamma)$ were obtained from the scaling collapse of the correlation function $C^o(r,\Gamma)$ (blue disks).
Second, as a comparison, we also plot the length scale $\xi_{c}(\Gamma)$ (orange crosses) 
at which the integrated correlation function falls below a given threshold:
 \begin{equation}
 \Sigma(\xi_c)=c    
 \label{eq:xic}
 \end{equation}
 for a threshold parameter $c \in (0,1)$~\cite{Holzner2009}. 
 Although simpler, this second method has the disadvantage of being sensitive to finite-size effects and convergence, as illustrated by the saturation of $\xi_{0.1}$ (orange crosses) for smaller couplings. These effects are reduced in the case of the scaling collapse approach as it takes into account the correlation function calculated at all odd sites.
 As a third method, we show the length scale $\xi_{K,S}(\Gamma)$ (green square) obtained from the scaling collapse of the impurity contribution to the entanglement entropy (near the impurity). This method will be  described in the following section.
 As scaling collapse methods determine the screening length only up to a global prefactor, we scale the data sets such that all three methods result in the same length scale for $1/\Gamma = 5$.
 
 In the Kondo regime ($\Gamma/U\ll 1$), we verify that the length scale follows the expected scaling
\begin{equation}
    \xi_K \sim \frac{1}{\sqrt{U\Gamma}} \exp\parO{\frac{\pi U}{8\Gamma}},
\end{equation}
where the prefactor is obtained through a fit (dashed black line) to the blue disks. In the regime of intermediate coupling strength all methods follow the expected exponential scaling. Away from this regime, in the case of weak coupling different level of sensitivity to finite size effect and convergence leads to underestimate the screening length compared to the expected exponential scaling. Similarly, in the strong coupling regime the mapping from the SIAM to the Kondo model is no longer valid and deviations are expected.

\subsection{Impurity entropy}\label{subsec:S2SIAM}
As a second probe of the SGS variational ground state, 
we consider the contribution of the Anderson impurity to the entanglement entropy.
This quantity offers a different approach to study the screening cloud of an impurity~\cite{Sorensen2007,Sorensen2007a,Affleck2009}. 

We consider a bipartition of the sites in subsystems A and B and compute the second R\'{e}nyi entropy
\begin{equation}
    S_2(A,\ket{\psi}) = - \log \Tr \parO{\rho_A^2},
\end{equation}
where $\rho_A = \Tr_B \ket{\psi}\bra{\psi}$ is the reduced density matrix for subsystem $A$. Throughout, subsystem A is formed of the $l$ first sites of the lead and the impurity.
While in the case of a single Gaussian state the entropy can easily be computed from the decomposition in normal modes of the covariance matrix~\cite{Botero2003,Vidal2003}, the presence of coherences between the different Gaussian states forming the ansatz in the density matrix modify this calculation.
In Appendix~\ref{App:S2} we use the fermionic coherent state formalism~\cite{Cahill1999} to derive expressions of $S_2(A,\ket{\psi})$ for an SGS state of arbitrary rank.

Even though $S_2$ differs from the more common von Neumann entropy $S_1$, many of the same universal properties can be extracted.
In particular, using conformal field theory (CFT) calculations, the constant contribution of the boundary to the entropy was shown to be independent of the order of the Rényi entropy at criticality (see e.g. Refs.~\cite{Zhou2006,Calabrese2009}). In addition the form of the leading corrections to the scaling of the entropy with subsystem size $l$ due to irrelevant boundary operators was shown to be independent of the order $n$ of the Rényi entropy~\cite{Eriksson2011,Eriksson2011a}.

As for the correlation function in Sec.~\ref{sec:spinspin}, for a lattice model at half-filling, the entanglement entropy is the sum of a uniform and a staggered contribution
\begin{equation}
    S_2(l,\Gamma,L)  = S_2^U(l,\Gamma,L) + (-1)^l S_2^A(l,\Gamma,L).
\end{equation}
The uniform and staggered contributions to the entropy can be extracted using a local polynomial interpolation~\cite{Sorensen2007a}. The impurity entropy $S_{\rm imp}(l,\Gamma, L)$ is then obtained by the subtraction 
\begin{equation}
    S_{\rm imp}(l,\Gamma, L) = S_2^U(l,\Gamma,L) - S_2^{0,U}(l,L)
\end{equation}
where $S_2^{0,U}(l,L)$ is the uniform part of the entropy in the absence of the impurity. 

\begin{figure}[tb]
    \centering
    \includegraphics[width=0.49\textwidth]{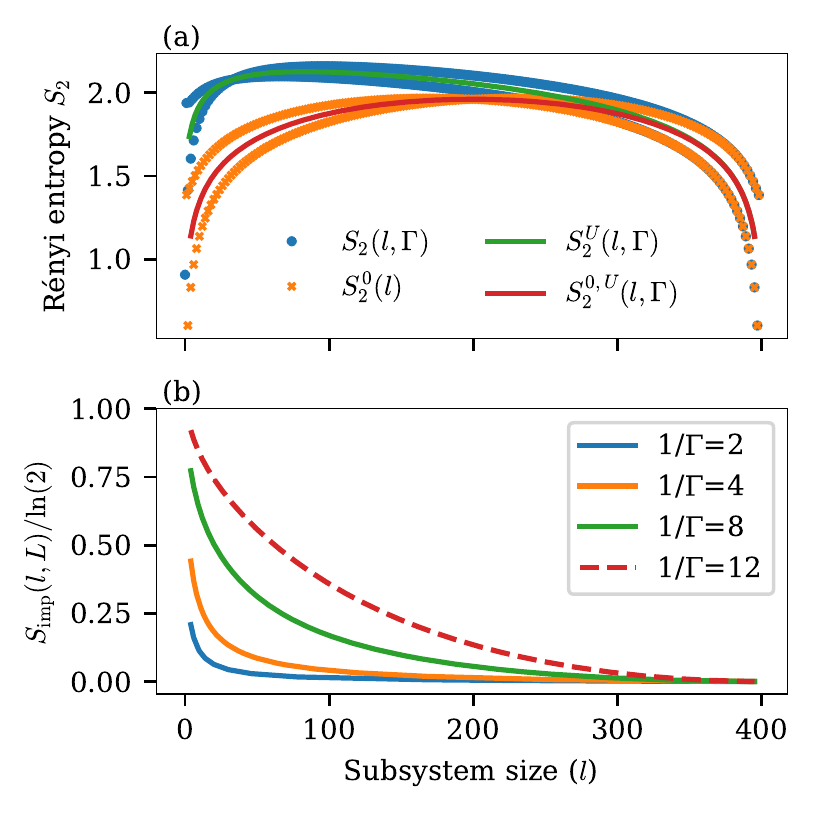}
    \includegraphics[width=0.49\textwidth]{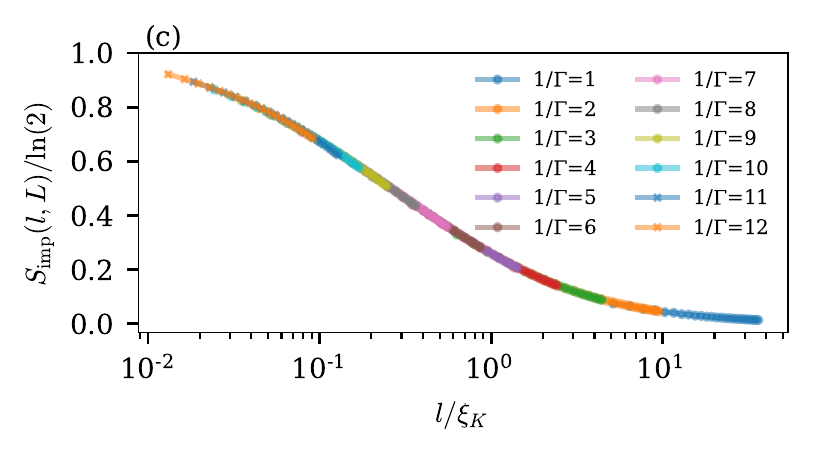}
    \caption{(a) 
    Raw data for the Rényi entropy $S_2(A)$ with (blue disks) and without (orange crosses) the impurity ($L=400$, $\Gamma=1/12$, $R=8$). Solid curves are the uniform part of the entropy extracted using a local polynomial interpolation~\cite{Sorensen2007a}.
        (b) Impurity entropy for different coupling strengths. Dashed red curve corresponds to  the subtraction of the green and red curves of panel (a). (c) Scaling collapse of $S_{\rm imp}$ close to the impurity. To reduce finite size effects, we consider only the first 25 sites. See Fig.~\ref{fig:K1}(c) for the extracted screening length $\xi_{K,S}(\Gamma)$.
    }
    \label{fig:K1_S2_L399}
\end{figure}
To make the calculation of $S_{\rm imp}$ more explicit, Fig.~\ref{fig:K1_S2_L399}(a) presents the raw data for a system of length $L=400$ and $\Gamma=1/12$. The corresponding impurity entropy $S_{\rm imp}$ is the dashed red curve in Fig.~\ref{fig:K1_S2_L399}(b).
As expected, for small couplings ($\Gamma\ll 1$) and close to the impurity ($l \ll \xi_K$), $S_{\rm imp} \sim \ln(2)$ indicating the entanglement of the impurity with the lead. Consistent with Fig.~\ref{fig:K1}(c), the characteristic length scale over which the $S_{\rm imp}$ decays increases as $\Gamma$ is reduced.

To extract more quantitatively the screening length $\xi_K$, we again consider a scaling collapse of the data. 
The impurity entropy was previously found to be amenable to scaling using~\cite{Sorensen2007}
\begin{equation}
    S_{\rm imp}(l, \Gamma, L) = S_{\rm imp}\left(\frac{l}{\xi_K(\Gamma)},\frac{l}{L}\right).
\end{equation}
Figure~\ref{fig:K1_S2_L399}(c) shows this scaling collapse where,
focusing on points near the impurity where ($l \leq 25$ and $L=400$), we consider $l/L \sim 0$. The correlation length extracted from the collapse $\xi_{K,S}(\Gamma)$ is shown in Fig.~\ref{fig:K1}(c) (green squares $\xi_{K,S}$). In the regime where the mapping from the SIAM to the Kondo model is valid ($U/\Gamma \gg 1$) and $\xi_K \ll L$, we obtain a good agreement between the different method to extract the Kondo screening length.

\section{Two-channel Kondo effect}\label{sec:MCK}
As a second application of the variational method, we consider the two-channel Kondo (2CK) impurity model.
We focus on the symmetric regime where the impurity is coupled with the same strength $J$ to both channels.
In this regime, the presence of a second channel of free electrons lead to very different 
ground state properties from the single channel case and the physics of the model includes an intermediate coupling fixed point with non-Fermi liquid behavior~\cite{NozieresPh.1980}. 

We focus on the ground state of the lattice version of the symmetric two-channel Kondo Hamiltonian
\begin{equation}
    H_{2CK} = -t \sum_{j,\alpha} \left( d^\dag_{j+1,\alpha}d_{j, \alpha} + h.c.
    \right)+H_J +H_{BC},
\end{equation}
where $\alpha =1,2$ is the channel index, $H_J$ is the coupling between the impurity spin and the channels of free fermions and $H_{BC}$ accounts for the choice of boundary condition away from the impurity.
We consider directly an antiferromagnetic spin-spin interaction of coupling strength $J>0$ between the impurity spin and the first site of each channel:
\begin{equation}
    H_J = J\sum_{\alpha=1,2} \mathbf{S}_{\rm imp} \cdot \mathbf{S}_\alpha(1),
\end{equation}
 where $ \mathbf{S}_\alpha(r) = \frac{1}{2}\sum_{s,s'} d^\dag_{r,\alpha, s} \boldsymbol{\sigma}_{s,s'} d_{r,\alpha, s'}$ is the fermionic spin operator.
As the SGS ansatz is fermionic, we take the impurity to be a fermionic site with fixed single-occupancy.\footnote{
As the charge of the fermionic impurity is conserved by $H_{2CK}$,
the ground state of this model with a fermionic impurity will be the same as in the case where the system is explicitly projected to the single-occupancy subspace (spin-$1/2$ impurity). We have verified numerically that charge fluctuations at the impurity site vanish in the variational ground state.
}
We also introduce the unitless coupling parameter
$g \equiv \rho_0 J$ with $\rho_0 = 1/4t$ the density of state at the Fermi Level.

For finite-size systems, the two-channel Kondo model is known to exhibit important differences between the case where the total number of sites is even or odd~\cite{Alkurtass2016}.
In order to preserve the symmetry between the two channels and have an even number of sites (including the impurity), we introduce an additional site at the opposite end of the chain coupling the two channels
\begin{equation}
    H_{BC} = -t \sum_{\alpha=1,2} \left(d_L^\dag d_{L-1, \alpha}  +h.c.
    \right),
\end{equation}
leading to a total of $2L$ sites in the model. 

\begin{figure}[tb]
    \centering
    \includegraphics[width=0.49\textwidth]{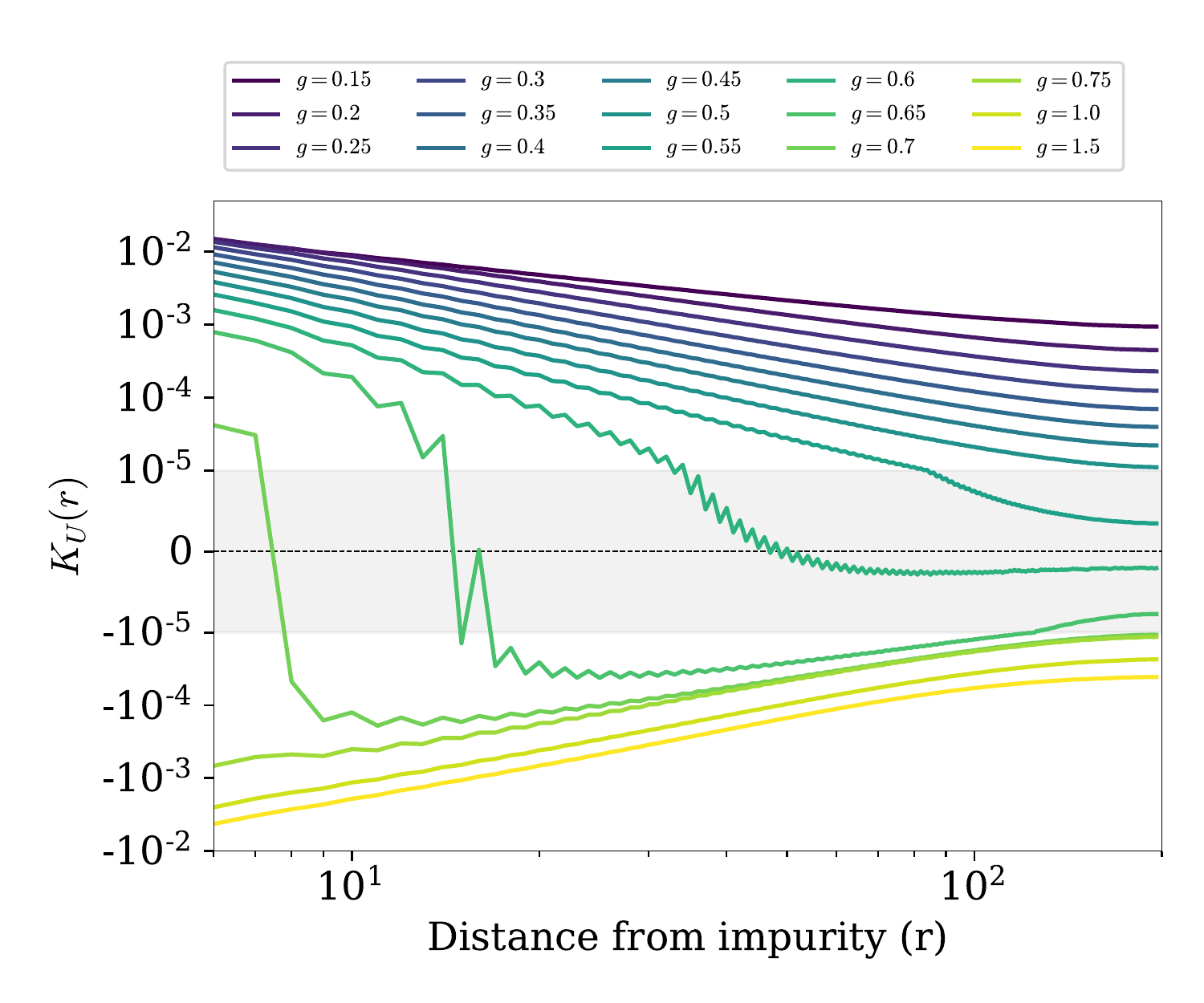}
    \caption{
    Uniform part of the spin-spin correlation function $K(r)$ (see Eq.~\eqref{eq:K2CK}) on a symmetric log axis. A sign change of the correlation function can be observed for $g\sim 0.7$ consistent with the crossing of the 2CK intermediate coupling fixed point (see maintext).
    Each channel is half-filled with hopping parameter $t=1/2$ and $L=200$ sites.
    An SGS variational state with rank $R\geq 25$ is used for all coupling strengths. The shaded area indicate the linear part of the symmetric logarithmic vertical axis.
    }
    \label{fig:KU2CK}
\end{figure}
\begin{figure}[tb]
    \centering
    \includegraphics[width=0.49\textwidth]{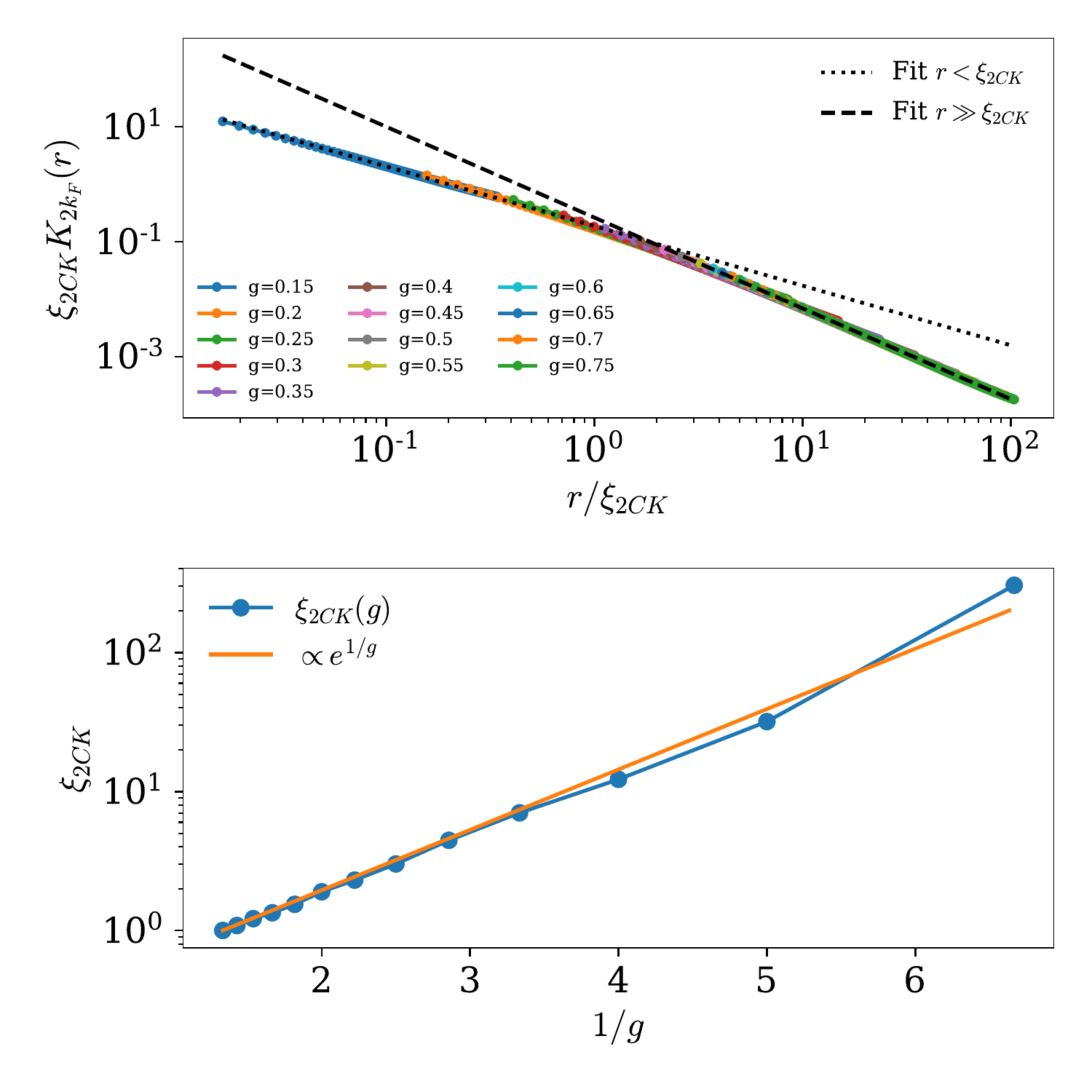}
    \caption{
    (a) Scaling collapse of $K_{2k_F}(r)$, the staggered part  of the spin-spin correlation function for various unitless coupling strengths $g=0.15 - 0.75$, with $L=200$ and, to limit finite-size effects, 
    $r\leq L/2$. 
    Black dashed (dotted) is a fit of the form $A(r/\xi_K)^{-\nu}$ to the data far (close) to the impurity with exponent $\nu  = 1.58$ (1.04).
    (b) Screening length $\xi_{2CK}$ extracted from the scaling collapse. Orange line is $\propto e^{1/g}$ with the proportionality factor set by scaling the first point of the curve.
    }
    \label{fig:corrxi2CK}
\end{figure}

\subsection{Intermediate coupling fixed point and screening cloud}
The 2CK model exhibits an intermediate coupling fixed point, as well as a duality between the weak and strong coupling regimes~\cite{Kolf2007}. 
In order to explore this physics in the SGS variational ground state, we consider the spin-spin correlation function
\begin{equation}
    K(r) = \sum_{\alpha = 1,2}\langle \mathbf{S}_{\rm imp} \cdot \mathbf{S}_\alpha (r) \rangle,
    \label{eq:K2CK}
\end{equation}
where $r$ is the distance to the impurity and $\alpha$ is the channel index.
We separate again the correlation function in a uniform and a staggered part
$K(r) = -K_U(r) - \cos(2 k_F r) K_{2k_F}(r)$.

Figure~\ref{fig:KU2CK} shows the uniform part of the correlation function $K_U(r)$ for various unitless coupling strengths $g=0.15-1.5$ on a symmetric logarithmic axis. Considering first $K_U$ close to the impurity, we observe a reduction of correlations as the coupling strength is increased. In absolute value, $|K_U(r)|$, reaches a minimum for $g^*\sim 0.7$ followed by a sign change of $K_U(r)$ near the impurity. We associate this point where correlations are minimal (smallest screening cloud) as the intermediate coupling strength fixed point of the 2CK model. This result is in agreement with previous NRG results for the Kondo temperature $T_{2CK}$~\cite{Kolf2007}.

In the weak coupling regime ($g\lesssim g^*$), the Kondo screening length of the 2CK model is to leading order $\xi_{2CK} \propto e^{1/g}$~\cite{Barzykin1998, Kolf2007}. 
In order to extract $\xi_{2CK}(g)$, we use a scaling collapse of $K(r)$.
We focus on the staggered part of the correlation function in the weak coupling regime where correlations are larger. To limit finite-size effects, we consider only correlations up to a distance $L/2$ from the impurity. 

Figure~\ref{fig:corrxi2CK}(a) presents a scaling collapse following the same  procedure as in Sec.~\ref{sec:anderson1}. For $g =0.15-0.75$, all correlation functions can be collapsed on a universal scaling function exhibiting a crossover between two different power laws near $r \sim \xi_{2CK}$. Using power law fits, we obtain the exponent $\nu = 1.04$ near the impurity and $\nu =1.58$ far from the impurity (dotted and black dashed lines). An exponent $\nu <2$ is coherent with the expected non-Fermi liquid behavior of an overscreened multichannel Kondo model. In particular,
for $r\gg \xi_{2CK}$ the scaling $K_{2k_F}(r) \propto (r/\xi_{2CK})^{1.5}$ was calculated in Ref.~\cite{Barzykin1998}.

The extracted screening length $\xi_{2CK}(g)$ is shown in Fig.~\ref{fig:corrxi2CK}(b) and compared to the expected exponential scaling (orange curve). Small deviations from the expected exponential behavior is expected for smaller values of $g$ where finite-size effects become important ($\xi_{2CK} \gtrsim L$).  
This result is also in agreement with previous DRMG calculations for an effective spin chain representation of the symmetric 2CK model where $\xi_{2CK}$ was extracted from a scaling collapse of the impurity entanglement entropy~\cite{Alkurtass2016}. 

\section{Conclusions}\label{sec:Conclusion}
In this work, we have developed a practical method for finding the variational ground state of an impurity problem using a coherent superposition of fermionic Gaussian states ansatz.
The approach has a computational complexity $O(R^2 N^3)$, thus scaling polynomially with the rank $R$ of the ansatz and the number of fermionic modes $N$. Combined with the favorable scaling of the accuracy with rank $R$, as guaranteed by the results of Ref.~\cite{Bravyi2017}, this gives a powerful new approach to study the ground state of quantum impurity models. In particular, the approach is independent of spatial locality or lattice connectivity, allowing for more flexibility than methods such as DMRG. In addition, its implementation is highly parallelizable allowing for further speed improvements.

In order to verify the method we have first studied the single impurity Anderson model. 
Comparing the ground state energy obtained using DMRG to the one obtained using the SGS ansatz, we have shown that an SGS state of rank $R\sim 6$ can rival with high precision DMRG calculations with a large bond dimension $m= 3000$. Further increasing the ansatz rank up to $R=16$ allows to improve the precision of the ground state energy estimate by an order of magnitude. To demonstrate the quality of the variational ground state, we have also carefully examined the spin-spin correlation functions and the impurity entanglement entropy, and found excellent agreement with analytical and previous numerical results.

To highlight the potential of the method, we have also studied the two-channel Kondo model.  
Studying again a spin-spin correlation function, we found signatures of the intermediate coupling fixed point in the form a sign change of the uniform part of the correlation function, and were able to confirm the expected exponential scaling of the Kondo screening length. We find a  power law decay of the correlation function far from the impurity with exponent $\nu \approx 1.58$, which is close to the expected non-Fermi liquid behavior with exponent $\nu =1.5$ predicted by CFT calculations~\cite{Barzykin1998}. These results showcase the power of the SGS ansatz for the study of real-space properties of multichannel impurity models, a space of applications previously ill-covered by current standard numerical methods.

An open problem is whether there are computationally efficient ways of extracting the impurity Green's function (in real or imaginary time or frequency) within this class of ansatz states. Such a method would allow this approach to be integrated as impurity solver in embedding methods, such as the dynamical mean-field theory, which requires the impurity Green's function to achieve self-consistency. Furthermore, it appears possible to further reduce the scaling of the method by replacing the Gaussian covariance matrices by Gaussian fermionic matrix product states~\cite{Schuch2019}, which could in principle further reduce the scaling with the number of fermionic modes $N$.

\acknowledgments{
    DMRG calculations were performed using the ITensor Library~\cite{ITensor}. Numerical calculations of pfaffians were performed using the PFAPACK package~\cite{Wimmer2012}.
    The authors thank A.W.W. Ludwig for insightful discussions.
}

\appendix

\section{Numerical solution of Eq.~\eqref{eq:EOM-Gamma3}}\label{app:numerical_details}
In this Appendix, we provide additional details on the numerical implementation of the variational method. There are two main approaches to solving Eq.~\eqref{eq:EOM-Gamma3} numerically. The first is to solve the equation using a generic differential equation solver, for example taking to first order
\begin{equation}
    \Gamma^\mu(s+\delta s) = \Gamma^\mu(s) + \delta s \del_s \Gamma^\mu(s),
    \label{eq:RKstep}
\end{equation}
with $\delta s$ the step size. 
The drawback of this approach is that the accumulation of small numerical errors due to the finite step size will rapidly lead to covariance matrices which do not represent normalized and pure states. One must then frequently correct the normalization by decomposing the covariance matrix in the canonical form
\begin{equation}
    \Gamma^\mu = R \bigoplus_{j=1}^N \matO{0 & \lambda_j \\ -\lambda_j &0 } R^T,
    \label{eq:CM-normal-form}
\end{equation}
with $R$ a real orthogonal matrix and rescaling the coefficients $\lambda_j$ to unity.

A second approach is to rewrite Eq.~\eqref{eq:RKstep} as an orthogonal transformation~\cite{Kraus_2010}
\begin{equation}
    \Gamma^\mu(s+\delta s) = R(s)\Gamma^\mu(s) R^T(s) + O(\delta s ^2),
    \label{eq:ExpStep}
\end{equation}
with the orthogonal matrix
\begin{equation}
    R(s) = \exp \parC{\frac{\delta s}{2} [\Gamma^\mu(s), \del_s \Gamma^\mu(s)]  },
\end{equation}
where we used $(\Gamma^\mu)^2=-\id$ and neglected terms quadratic in $\delta s$. This second approach is numerically more expensive than the former due to the required matrix exponentiation. However, it preserves normalization up to machine precision.
Our numerical experiments showed this second approach to perform better in some cases as it allows for taking larger time steps and one need not frequently correct the normalization of the covariance matrices  using costly canonical form decomposition.

\section{Entanglement entropy of SGS states}\label{App:S2}
In this Appendix, we derive the expression used to compute the order 2 Rényi entropy $S_2(\ket{\psi},A)$ in Sec.~\ref{subsec:S2SIAM}.  We consider a bipartition of an SGS state $\ket{\psi}$ in parts A and B. 
Defining the density matrices  $\rho_\alpha = \ket{\phi_\alpha} \bra{\phi_\alpha}$, the reduced density matrix for subsystem $A$ is
\begin{align}
    \rho_A &= \Tr_B(\ket{\psi} \bra{\psi} ) 
    =  \sum_{\alpha, \beta} \lambda_\alpha \lambda_\beta^* 
    \frac{\Tr_B\parS{\rho_\alpha \rho_\beta}}{G_{\alpha ,\beta}},
    \label{eq:rhoab}
\end{align}
since the states in the SGS ansatz are non-orthogonal
and thus $G_{\alpha, \beta}=  \braket{\phi_\alpha}{\phi_\beta} \neq 0$.
In order to evaluate Eq.~\eqref{eq:rhoab}, we expand the density matrix in a fermionic coherent state basis~\cite{Cahill1999}.
The derivation is sketched below with the final result given by Eq.~\eqref{eq:trB_rhoAB}. 

Using Eq.~\eqref{eq:rhoab}, the Rényi entropy of order 2 of an SGS state
is 
\begin{equation}
    S_2(\ket{\psi},A) = -\ln\parC{
        \sum_{\alpha,\beta, \gamma, \delta} 
        T_{\gamma, \delta}^{\alpha,\beta}
    },
    \label{eq:S2_SGS}
\end{equation}
where we introduced the rank-4 tensor
\begin{equation}
    T_{\gamma, \delta}^{\alpha,\beta} = 
    \frac{\lambda_\alpha \lambda_\beta^* \lambda_\gamma \lambda_\delta^*}{G_{\alpha, \beta}G_{\gamma, \delta}}
    I_{\gamma,\delta}^{\alpha,\beta}
        ,
        \label{eq:T_tensor}
\end{equation}
which obeys the relations
    $T_{\gamma, \delta}^{\alpha,\beta} = (T^{\beta, \alpha}_{\delta,\gamma})^*$ and 
    $T_{\gamma, \delta}^{\alpha,\beta} =  T^{\gamma, \delta}_{\alpha,\beta}$ and where 
    $I_{\gamma,\delta}^{\alpha,\beta}=\Tr_A(\Tr_B(\rho_\alpha \rho_\beta)\Tr_B(\rho_\gamma \rho_\delta ))$
    .
The final result, as a function of the covariance matrices of the Gaussian states, is given by inserting Eq.~\eqref{eq:TrArho2_final} in the above equation.
    
\subsection{Coherent state operator expansion}
Setting first the required notation, we introduce for each fermionic mode the coherent states
\begin{equation}
    \ket{\eta}_i = (1- \eta_i \hat d_i^\dag)\ket{0}_i, 
    \qquad 
    \bra{\etab}_i = \bra{0}_i(1 - \hat d_i \etab_i) ,
\end{equation}
where $\eta_i, \etab_i$ ($i=1, \dots N$) are Grassman variables 
obeying the usual algebra
\begin{equation}
    \begin{split}
    \eta_i^2 &=\etab_i^2 =0\\
    \eta_i \etab_j &= -\etab_j \eta_i,\\
    \eta_i \eta_j &= -\eta_j \eta_i.
    \end{split}
\end{equation}
To lighten the notation we introduce the states
$\ket{\eta} = \bigotimes_{i} \ket{\eta}_i$ and $\bra{\etab} = \bigotimes_{i} \bra{\etab}_i$, as well as the shorthands $\etab \cdot \eta =\sum_i \etab_i \eta_i$ for products and 
$d^{N}\eta = \prod_{j=1}^N d\etab_j d\eta_j$ for differentials. In addition, we make the dependency on barred variables implicit such that $F(\eta,\etab) \ra F(\eta)$.

Following the results of Ref.~\cite{Cahill1999}, any operator $\Om$ can be represented by the integral
\begin{equation}
    \Om
    = \int d^{N} \eta \,\, \chi(\Om,\eta) F(\eta)
    \label{eq:OpExp},
\end{equation}	
where 
$\chi(\Om,\eta)$
is the characteristic function of the operator
\begin{align}
    \chi(\Om,\eta)
    = \Tr\parS{\Om D(\eta)},
    \label{eq:characteristicFct}
\end{align}
and the operator $F(\eta)$ is
\begin{equation}
    F(\eta) = \int d^{N}\psi \, \, e^{\frac{1}{2}\etab \cdot \eta-\psib \cdot\psi 
    +\psi \cdot \etab - \eta \cdot \psib
    }
    \ket{\psi} \bra{-\psib}.
    \label{eq:Fop}
\end{equation}
Following the language of quantum optics $D(\eta) = \exp\parC{d^\dag \cdot \eta - \etab \cdot d} $ is the fermionic analogue of the bosonic displacement operator.

In this work, the operator $\Om$ in Eq.~\eqref{eq:OpExp} is the density matrix of a fermionic 
Gaussian state~\cite{Bravyi2004,Botero2003}
\begin{equation}
    \rho = \frac{1}{2^N} \prod_{j=1}^N \parO{\id  - i \lambda_j b_{2j-1} b_{2j}},
\end{equation}
with the canonical modes $b_j = \sum_i R_{i,j}c_i$ with $R \in SO(2N)$.  The rotation matrix $R$ and the eigenvalues $\lambda_j$
 are defined through the normal form decomposition of the covariance matrix introduced in Eq.~\eqref{eq:CM-normal-form}.
Using this decomposition and rewriting the displacement operator in this canonical basis, the trace in Eq.~\eqref{eq:characteristicFct} can be evaluated mode per mode leading to the characteristic function
\begin{equation}
    \chi(\rho,\eta) = \exp\parC{
         \frac{-i}{2} 
         \matO{\etab^T & \eta^T } V_N^T\Gamma V_N \matO{\etab \\ \eta}
        },
    \label{eq:chiRho}
\end{equation}
where the $2N \times 2N$ matrix $V_N$ makes the rotation of the Grassman operators from a complex fermion basis to a real Majorana basis:
\begin{equation}
    V_1 \matO{\etab_1 \\ \eta_1} = \frac{1}{2}\matO{\etab_1 + \eta_1 \\ i (\etab_1 - \eta_1)}.
    \label{eq:basisChange}
\end{equation}
The characteristic function Eq.~\eqref{eq:chiRho} allows to relates the Grassman variable representation used e.g. in Refs.~\cite{Bravyi2004,Bravyi2017} 
to expressions involving the density matrix through Eq.~\eqref{eq:OpExp}.

\subsection{Reduced density matrix}
We now turn to deriving the necessary expressions for evaluating numerically Eq.~\eqref{eq:rhoab}.
To this end, we need to compute partial traces of the form $\Tr_B(\rho_\alpha \rho_\beta)$, where $\Tr_B$ is the partical trace over subsystem $B$ which is composed of $N_B$ modes such that $N=N_A+N_B$.

Denoting states and Grassman variables associated with the subsystem by a subscript, the partial trace in the coherent state basis takes the form
\begin{equation}
    \Tr_B(\rho_\alpha \rho_\beta) = \int d^{N_B}\psi_B  \, e^{-\psib_B \cdot \psi_B} \bra{-\psib_B} \rho_\alpha \rho_\beta \ket{\psi_B}.
    \label{eq:tr}
\end{equation}
\begin{widetext}
Using Eq.~\eqref{eq:OpExp}
and noting that the $F$ operator is separable such that $F(\eta) = F(\eta_A)F(\eta_B)$, we obtain 
\begin{align}
    \Tr_B (\rho_\alpha \rho_\beta) &=
        \int d^{N}\eta \,\, \chi(\rho_\alpha,\eta) F(\eta_A)
        \int d^{N} \lambda \,\, \chi(\rho_\beta,\lambda) F(\lambda_A)
    K(\eta_B,\lambda_B)
    \label{eq:trb1}
\end{align}
where we introduced the kernel
\begin{align}
    K(\eta_B,\lambda_B) 
    &= \int d^{N_B}\psi_B \,\, e^{-\psib_B \cdot \psi_B}\, \bra{-\psib_B} 
    F(\eta_B)F(\lambda_B)
    \ket{\psi_B}.
\end{align}
Using Eq.~\eqref{eq:Fop} and evaluating the resulting Gaussian integral we obtain
\begin{align}
    K(\eta_B,\lambda_B)  &= 
    2^{N_B}
    \exp\parC{\frac{1}{2}(\etab_B \cdot \lambda_B - \lambb_B \cdot \eta_B},
    \label{eq:Kres}
\end{align}
where we used for example $\braket{-\psib}{\eta} = \exp(-\psib \eta)$.

Inserting Eq.~\eqref{eq:Kres} in Eq.~\eqref{eq:trb1}, we now isolate the integrals over subsystem $B$ 
\begin{equation}
    \Tr_B (\rho_\alpha \rho_\beta) =
        \int d^{N_A}\eta_A  d^{N_A} \lambda_A \,\,
        F(\eta_A)
        F(\lambda_A)
        \Xi_{\alpha,\beta}(\eta_A,\lambda_A)
        \label{eq:trB_rhoAB}
\end{equation}
with the effective characteristic function for the subsystem $A$
\begin{equation}
    \Xi_{\alpha,\beta}(\eta_A,\lambda_A) = 
    \int d^{N_B}\eta_B  d^{N_B} \lambda_B \,\,
        \chi(\rho_\alpha,\eta)
        \chi(\rho_\beta,\lambda) 
        K(\eta_B,\lambda_B).
        \label{eq:Xi1}
\end{equation}
In order to take advantage of the structure of Eq.~\eqref{eq:basisChange},
we formalize the basis change of Eq.~\eqref{eq:basisChange} by introducing the new Grassman variables $\theta_i  = \sum_j (V_N)_{i,j}(\etab, \eta)_j$
with the differentials transforming as
$\drm \etab_j \drm \eta_j  = \frac{-i}{2} \drm \theta_{2j-1}\drm \theta_{2j}$.
Using $\tilde{(\cdot)}$ to indicate functions with arguments in the rotated basis, Eq.~\eqref{eq:Xi1} takes the form
\begin{equation}
    \tilde{\Xi}_{\alpha,\beta}(\theta_A, \phi_A) =
    \parO{\frac{-1}{2}}^{N_B}\int D\theta_B 
        D\phi_B \,\,
        \tilde{\chi}(\rho_\alpha,\theta)
        \tilde{\chi}(\rho_\beta,\phi) 
        e^{\theta_B^T \cdot \phi_B}
        \label{eq:Xi1}
\end{equation}
where $D\theta_B = \prod_{j=2N_A+1}^{2N} d\theta_j$. In the case $N_B=N$, i.e. when tracing over the whole system, we recover the result of Ref.~\cite{Bravyi2004}.

Introducing the block structure of the covariance matrix
\begin{equation}
    \Gamma^\mu  = \matO{\Gamma^{\mu}_A & \Gamma^{\mu}_{AB} \\ \Gamma^{\mu}_{BA} & \Gamma^{\mu}_{B}}
\end{equation}
where $\Gamma_A$, $\Gamma_B$ are skewsymetric and $\Gamma_{AB} = - \Gamma_{BA}^T$, one can evaluate the Gaussian integrals resulting in 
\begin{equation}
    \tilde{\Xi}_{\alpha,\beta}(\theta_A, \phi_A) =
    2^{-N_B}
    \Pf\parS{
        M_B^{(\alpha,\beta)}
    }
    \exp\parC{
        -\frac{i}{2}\matO{  \theta_A^T 
        &  \phi_A^T}
        \parS{\matO{\Gamma_{A}^{\alpha} & 0 \\ 0 &\Gamma_{A}^{\beta }}+
        \Sigma_A^{\alpha,\beta}}
        \matO{  \theta_A
        \\   \phi_A}
    }
    \label{eq:XiFinal}
\end{equation}
where we introduced the $4 N_J \times 4 N_J$  skewsymmetric matrices ($J \in \{A,B\}$)
\begin{align}
    M_J^{\alpha,\beta} = \matO{\Gamma^{\alpha}_J & i\id\\ -i\id&\Gamma^{\beta}_J},
    \label{eq:MB}
    \qquad
            \Sigma_A^{\alpha,\beta} = 
        \matO{\Gamma_{AB}^{\alpha} & 0 \\ 0 &\Gamma_{AB}^{\beta }}
        \parS{M_B^{\alpha,\beta}}^{-1}
        \matO{\Gamma_{AB}^{\alpha} & 0 \\ 0 &\Gamma_{AB}^{\beta }}^T.
\end{align}
Finally, one can transform back to the original basis to obtain $\Xi(\eta_A,\etab_A,\lambda_A,\lambb_A)$, by taking $\theta_A = V_N (\etab_A, \eta_A)$ and $\phi_A = V_N (\lambb_A, \lambda_A)$.

\subsection{Trace of product of reduced density matrices}
Building on the results of the previous section, we now turn to the calculation of the trace needed to evaluate Eq.~\eqref{eq:T_tensor}. Dropping the subscript $A$ when there is no confusion and using Eq.~\eqref{eq:trB_rhoAB} the trace of the product of reduced density matrices takes the form
\begin{align}
    I_{\gamma,\delta}^{\alpha,\beta}
    &=
    \int d^{N_A}\eta 
    d^{N_A} \lambda 
    d^{N_A}\xi 
    d^{N_A} \nu 
    \,\,
    \Xi_{\alpha,\beta}(\eta,\lambda)
    \Xi_{\gamma,\delta}(\xi,\nu)
    \mathcal{K}(\eta,\lambda,\xi,\nu)
\end{align}
where we introduce the function
\begin{equation}
    \mathcal{K}(\eta,\lambda,\xi,\nu)=
    \int d^{N_A}\psi \,\,
    e^{-\psib \cdot \psi} 
    \bra{-\psib} 
    F(\eta)
    F(\lambda)
    F(\xi)
    F(\nu)
    \ket{\psi}.
\end{equation}
Inserting the definition of $F$ and performing five Gaussian integrals, one obtains
\begin{equation}
    \mathcal{K}(\eta,\lambda,\xi,\nu)=
    2^{N_A} 
    e^{\frac{1}{2}(\etab \cdot \lambda - \lambb \cdot \eta)}
    e^{\frac{1}{2}(\overline{\xi} \cdot \nu - \overline{\nu} \cdot \xi)}
    e^{\frac{1}{2}(\etab \cdot \nu - \overline{\nu} \cdot \eta)}
    e^{\frac{1}{2}(\lambb \cdot \xi - \overline{\xi} \cdot \lambda)}
    e^{\frac{1}{2}(\overline{\xi} \cdot \eta - \etab \cdot \xi)}
    e^{\frac{1}{2}(\overline{\nu} \cdot \lambda - \lambb \cdot \nu)}.
    \label{eq:Km}
\end{equation}
As all the coherent states have been eliminated, we rotate back to the real Majorana basis introduced in Eq.~\eqref{eq:Xi1} such that
\begin{equation}
    I_{\gamma,\delta}^{\alpha,\beta} 
    = 
    2^{-4N_A}
    \int 
    D\theta
    D\phi
    D\theta'
    D\phi'
    \,\,
    \tilde{\Xi}_{\alpha,\beta}(\theta, \phi)
    \tilde{\Xi}_{\gamma,\delta}(\theta', \phi')
    \tilde{\mathcal{K}}(\theta,\phi, \theta', \phi').
\end{equation}
Grouping the variables such that $\Theta = (\theta,\phi)$, $\Phi=(\theta', \phi')$ and inserting Eqs.~\eqref{eq:XiFinal} and \eqref{eq:Km}
\begin{equation}
    I_{\gamma,\delta}^{\alpha,\beta} 
    = 
    \frac{2^{N_B} }{8^{N}} \Pf[M_B^{\alpha,\beta}]\Pf[M_B^{\gamma,\delta}]
    \int D\Theta D\Phi \,\, 
    e^{\Theta^T X \Phi}
    \exp \parC{\frac{-i}{2} \parS{\Theta^T (M_A^{\alpha,\beta} +\Sigma_A^{\alpha,\beta})\Theta
    +\Phi^T (M_A^{\gamma, \delta} +\Sigma_A^{\gamma, \delta})\Phi
    }
    }
\end{equation}
where we introduced the matrix which couple the $\Theta$ and $\Phi$ varibales
\begin{equation}
    X = \matO{ -\id_{2N_A}& \id_{2N_A} \\\id_{2N_A} &-\id_{2N_A}}.
\end{equation}
Performing the final Gaussian integrals, we obtain
\begin{equation}
    I_{\gamma,\delta}^{\alpha,\beta} 
    = 
    \frac{2^{N_B} }{8^{N}} \Pf[M_B^{\alpha,\beta}]\Pf[M_B^{\gamma,\delta}]
    \Pf[M_A^{\gamma, \delta} +\Sigma_A^{\gamma, \delta}]
    \Pf[M_A^{\alpha,\beta} +\Sigma_A^{\alpha,\beta}+\sigma_A^{\delta,\gamma}],
    \label{eq:TrArho2_final}
\end{equation}
with the matrix
\begin{equation}
    \sigma_A^{\delta,\gamma} = -X\parO{
        M_A^{\gamma, \delta} +\Sigma_A^{\gamma, \delta}
    }^{-1} X.
\end{equation}
We note that Eq.~\eqref{eq:TrArho2_final} is symmetric under the exhange of indices 
$(\alpha,\beta) \leftrightarrow (\gamma, \delta)$ as required for a trace. This can be more explicit by considering the expression for the Pfaffian of a matrix with a block structure
\begin{equation}
    \Pf[M_A^{\gamma, \delta} +\Sigma_A^{\gamma, \delta}]
    \Pf[M_A^{\alpha,\beta} +\Sigma_A^{\alpha,\beta}+\sigma_A^{\delta,\gamma}]=
    \Pf \parS{
        \matO{
            M_A^{\alpha,\beta} +\Sigma_A^{\alpha,\beta} & iX  \\ -iX  & M_A^{\gamma, \delta} +\Sigma_A^{\gamma, \delta}
        }
    }.
    \label{eq:pfMA}
\end{equation} 
Although numerically more costly, the RHS of Eq.~\eqref{eq:pfMA} presents the advantage of making no assumption about the existence of the matrix inverse $\parO{
    M_A^{\gamma, \delta} +\Sigma_A^{\gamma, \delta}
}^{-1}$.
Together with Eq.~\eqref{eq:S2_SGS}, Eq.~\eqref{eq:TrArho2_final} allows to compute the order 2 Rényi entropy of an SGS state of arbitrary rank.

\end{widetext}

\bibliography{biblio_SGS}

\end{document}